\documentclass{article}

\usepackage[preprint]{neurips_2024}
\usepackage[utf8]{inputenc} 
\usepackage[T1]{fontenc}    
\usepackage{hyperref}       
\usepackage{url}            
\usepackage{booktabs}       
\usepackage{amsfonts}       
\usepackage{nicefrac}       
\usepackage{microtype}      
\usepackage{xcolor}         
\usepackage{caption, subcaption}
\usepackage{graphicx}

\title{Predicting Individual Depression Symptoms from Acoustic Features During Speech}

\author{%
  Sebastian Rodriguez \\
  Dalhousie University and Vector Institute\\
  Canada \\
  \texttt{sebastian@dal.ca} \\
  \And
  Sri Harsha Dumpala \\
  Dalhousie University and Vector Institute\\
  Canada \\
  \texttt{sriharsha.d@dal.ca} \\
  \AND
  Katerina Dikaios \\
  Dalhousie University and Nova Scotia Health\\
  Canada \\
  \texttt{katerina.dikaios@dal.ca} \\
  \And
  Sheri Rempel \\
  Nova Scotia Health\\
  Canada \\
  \texttt{sheri.rempel@nshealth.ca} \\
  \And
  Rudolf Uher \\
  Dalhousie University and Nova Scotia Health\\
  Canada \\
  \texttt{uher@dal.ca} \\
  \And
  Sageev Oore \\
  Dalhousie University and Vector Institute\\
  Canada \\
  \texttt{sageev@dal.ca} \\
}

\begin{document}
\maketitle

\begin{abstract}
 Current automatic depression detection systems provide predictions directly without relying on the individual symptoms/items of depression as denoted in the clinical depression rating scales. In contrast, clinicians assess each item in the depression rating scale in a clinical setting, thus implicitly providing a more detailed rationale for a depression diagnosis. In this work, we make a first step towards using the acoustic features of speech to predict \textit{individual items of the depression rating scale} before obtaining the final depression prediction. For this, we use convolutional (CNN) and recurrent (long short-term memory (LSTM)) neural networks. We consider different approaches to learning the temporal context of speech. Further, we analyze two variants of voting schemes for individual item prediction and depression detection. We also include an animated visualization that shows an example of item prediction over time as the speech progresses.
\end{abstract}

\section{Introduction}
\label{sec:intro}

Depression has been coined as a disease of modernity due to its increasing prevalence in recent times \cite{hidaka2012depression}. Multiple researchers have designed systems to detect its presence and severity using audio and video recordings \cite{al2018detecting, nature_sc_reports_2023, huang2020exploiting}. So far, the majority of approaches focus on predicting depression using the aggregate score of existing depression scales such as the Montgomery-Asberg Depression Rating Scale (MADRS) \cite{montgomery1979madrs} or the Patient Health Questionnaire 9 (PHQ-8) \cite{phq8}. A small number of works have analyzed the relation (correlation) between acoustic features and individual items/questions of the depression rating scales \cite{horwitz2013importance, trevino2011phonologically}. However, none of the previous studies have aimed to predict individual items of depression rating scales. Yet, predicting these individual items could provide a better understanding of the decision made by the machine learning models.

\begin{figure}[!htbp]
     \begin{subfigure}[b]{0.57\textwidth}
         \centering
         \includegraphics[width=\linewidth]{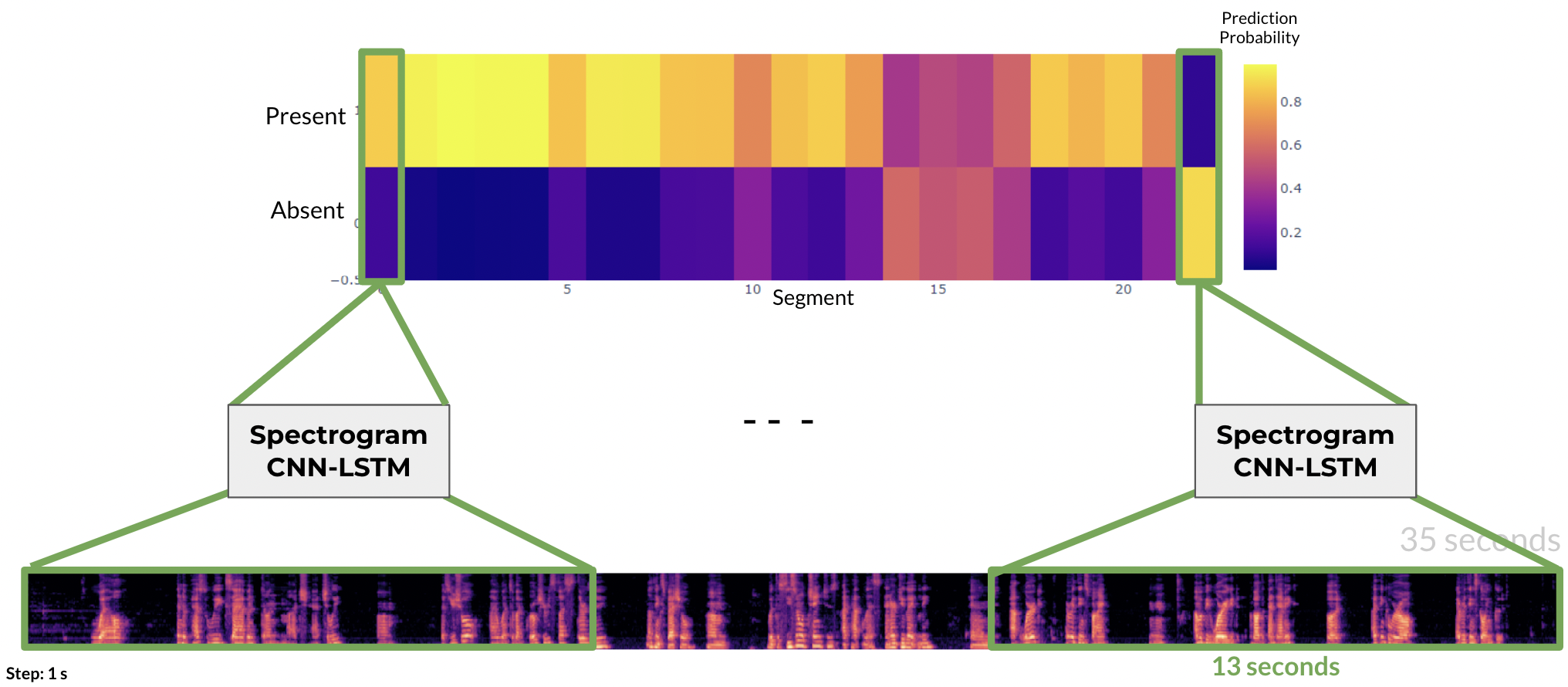}
         \caption{Processing speech as a time-series for depression assessment}
         \label{fig:timeseries_demo}
     \end{subfigure}
     \hfill
     \begin{subfigure}[b]{0.41\textwidth}
         \centering
         \includegraphics[width=\textwidth]{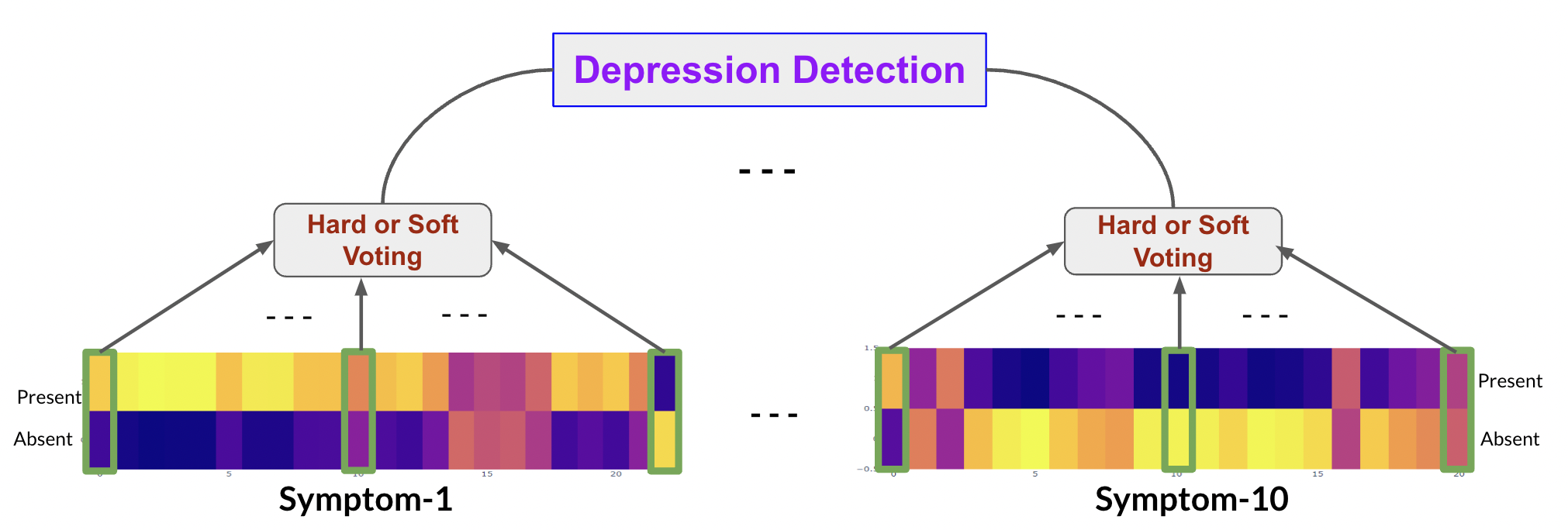}
         \caption{Final decision making using hard or soft voting}
         \label{fig:hard_soft_voting}
     \end{subfigure}
        \caption{Outline of our approach exploiting the temporal information in speech for depression assessment. (a) shows the output of model trained on  speech recordings provided as a time-series of short segments (input speech segmented into 13 second segments with an overlap of 1 second) for detecting suicidal thoughts -- one of the symptoms of major depressive disorder, (b) The segment level predictions obtained for each item are combined either using hard or soft voting. The individual item predictions are combined to obtain a final decision on depression prediction.}
        \label{fig:outline}
\end{figure}

In this paper, we analyze \textit{acoustic features}, extracted at segment-level, to estimate \textit{individual items of the depression rating scales}. 
An outline of our two-stage approach for depression detection is shown in Figure \ref{fig:outline}. In stage-1, as shown in Figure \ref{fig:timeseries_demo},  given a spectrogram represented as a time-series of short segments (input speech segmented into overlapping 13-second segments) as input, the model generates the probability of detecting an item (e.g. suicidal thoughts -- one of the symptoms of major depressive disorder). The segment-level probabilities are shown as a grid with 2 rows and 22 columns of cells. The 22 columns together represent a 35-second recording of speech, which has been split into 22 overlapping segments, each segment corresponding to a 13-second excerpt. This grid shows the predicted probability, for each segment, that a particular symptom is present based on the recorded speech. For example, segment 0, the leftmost second, is coloured bright yellow (upper) and dark purple (lower), meaning the  probability of suicidal thoughts being present at a level consistent with depression is 0.85 (and a corresponding probability of 0.15 that these thoughts are not present at this level). In the rightmost cell, the probabilities are approximately 0.1 and 0.9, indicating that that symptom is predicted likely to be absent based on the last 13 seconds of recorded speech in this 35-second excerpt. In stage-2, as shown in Figure \ref{fig:hard_soft_voting}, the segment-level probabilities of each item are combined---using either hard or soft voting---to obtain the final prediction for depression detection for the entire 35-second recording.

In this work, we find that learning local contextual information using CNN-LSTM (combination of convolutional neural network (CNN) and long-short term memory (LSTM)) network performs better than a CNN network on most of the items. Note that we are using acoustic features alone to predict symptoms whose presence was established in clinical interviews (see Section~\ref{s:datasets}); so for example, if the system predicts suicidal thoughts, then that means that based on the sound of the 35-second speech excerpt, it predicts that suicidal ideation was noted at some point during the entire interview, but generally it is \textit{not} the case that it is mentioned in any way during that particular 35-second excerpt.

To see a visualization of the predictions while listening to the corresponding section of speech, please see the videos attached in this {\color{blue}{\href{https://drive.google.com/drive/folders/18MdBxFF_iEK-7J9PpS4dyZVZFwLDk4j5?usp=sharing}{link}}}. Note that to protect privacy of patients, the speech in the attached video is simulated by an actor, while the speech used in the experiments described here was collected as described in Section~\ref{s:datasets}.

\section{Related Work}
In recent years, several models were proposed for the tasks of automated depression detection and severity estimation tasks. These solutions range from suggesting innovative features \cite{williamson2013vocal, huang2020exploiting} to exploring different neural network architectures \cite{ma2016depaudionet}. Moreover, data comes in different formats, leading to solutions that use text \cite{tao2020spotting} and video recordings \cite{yang2016decision}, in addition to audio.

When it comes to speech, neural architectures have been proposed using CNNs \cite{palaz2013estimating, dubagunta2019learning} and LSTMs \cite{al2018detecting}. Other works have used existing solutions from different domains to enhance the performance of their models \cite{dumpala2021estimating, zhao2019automatic, dumpala2021sine}.

Individual items from depression assessment frameworks have been used as a core element in depression biomarker studies \cite{trevino2011phonologically, horwitz2013importance, arsenievkoehler2018happiness}. To the best of our knowledge, no systems have been proposed to predict the individual items of depression rating scales.

\section{Datasets}
\label{s:datasets}

In this work, we used two depression datasets: (1) Distress Analysis Interview Corpus - Wizard of
Oz (DAIC-WOZ) \cite{gratch2014distress} and (2) Autobiographical Adult Speech Samples (AASS) for analysis.
DAIC-WOZ dataset contains a set of 219 clinical interviews collected from 219 participants.
Each audio sample was labeled with a Patient Health Questionnaire-8 (PHQ-8) score \cite{phq8}. PHQ-8 consists of 8 different items to rate the severity of depression. Each item in PHQ-8 is rated in the range of 0-3.
AASS dataset contains 131 speech samples collected from 109 participants. Depression severity of each speech sample was scored on the Montgomery and Asberg Depression Rating Scale (MADRS). MADRS consists of 10 different items to rate the severity of depression. Each item in MADRS is rated in the range of 0-6. In this study, if the score for an item is zero (Item-score = 0) --> the item is absent, and score is greater than zero (Item-score > 0)  --> item is present.
Further details are provided in the appendix.

\begin{figure}[!tbp]
  \begin{minipage}[b]{0.55\textwidth}
    \centerline{\includegraphics[width=1\linewidth]{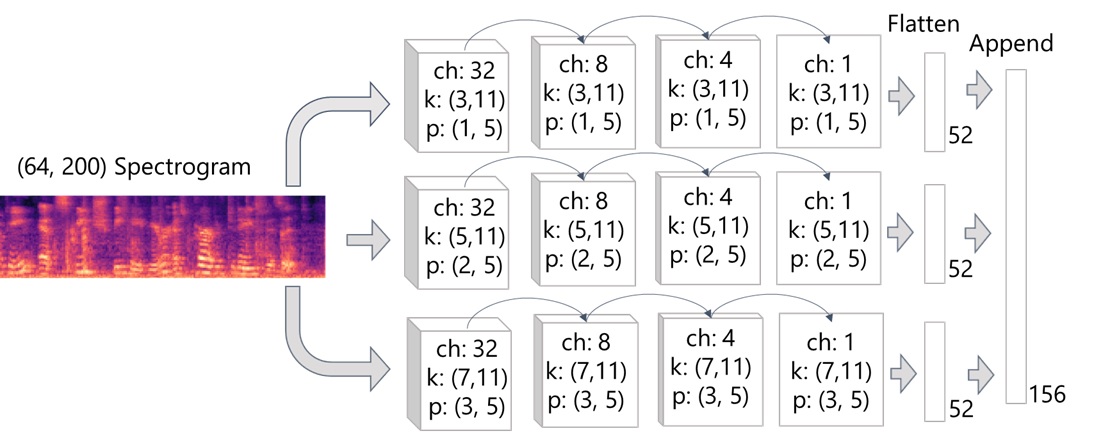}}
    \caption{Each box represents a 2-d convolution followed by a ReLU activation function. All the convolutions had a stride of (2, 2). ch: Number of channels, k: kernel/filter size, p: padding.}
    \label{fig:cnn}
    \end{minipage}
    \hfill
    \begin{minipage}[b]{0.435\textwidth}
    \centerline{\includegraphics[width=1\linewidth]{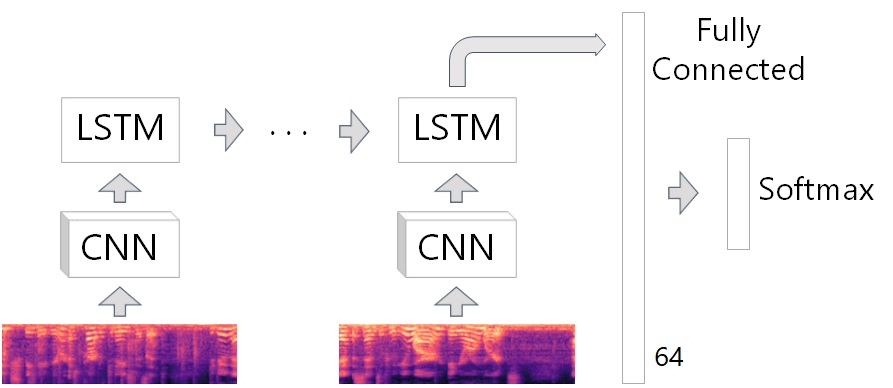}}
    \caption{Complete architecture of the Spectrogram CNN-LSTM model. The CNN corresponds to the convolutional component explained in Figure \ref{fig:cnn}. The LSTM layer has a hidden state size of 64.}
\label{fig:model}
  \end{minipage}
\end{figure}

\subsection{Features} 

Two different types of features are used, Spectrograms and eGeMAPS. Spectrograms are obtained
from the cleaned audio files by looking at 20 milliseconds (ms) windows, taking steps of 20 ms, and
64 filter banks are used to calculate the log Mel Spectrogram used in the experiments. Additionally,
the eGeMAPS feature set \cite{eyben2015egemaps} is extracted using the OpenSMILE toolkit \cite{eyben2010opensmile}. It is important to note that the eGeMAPS features are obtained by computing various statistics of the frame-level features.

\section{Models}

\label{section:models}
\textbf{Spectrogram CNN Model:} The main component of this network is described in Figure \ref{fig:cnn}, it consists of 3 parallel strided 2-dimensional convolutional layers, with different kernel sizes. The outputs of the layers are flattened and appended into a single 156-dimensional embedding. This embedding is sent through a linear layer that transforms the embedding into the desired binary logits. Finally, the softmax activation function is used over the outputs. In this model, each sample represents a spectrogram extracted from a 4-second speech segment.
\\
\textbf{Spectrogram CNN-LSTM Model:} This model  builds up on the Spectrogram CNN. The main component from Figure \ref{fig:cnn} is used across multiple spectrograms, the embeddings obtained are used as the input of an LSTM with a hidden state of 64, as explained in \ref{fig:model}. The output of the LSTM is sent through a fully connected layer that produces the logits received by the last activation activation function (softmax or linear). A total of 10 4-second spectrograms represent one sample. The spectrograms correspond to contiguous sections of an audio recording, taking a step of 1 second between one another. With this configuration one sample of the proposed model contains 10 spectrograms that correspond to a segment of 13 seconds of a recording. After the convolutional section of the model the embeddings are used as the input of a LSTM layer.

\section{Training}
Adam optimizer (parameters $\beta_1$:$0.9$, $\beta_2$: $0.999$, $\alpha$: $0.0005$) was used, along with $L2$ regularization, to minimize the Negative Log-likelihood loss function for the classification tasks and Mean Squared Error for the regression tasks. Additionally, a random search of hyper parameters was applied for deciding on the usage of batch normalization and dropout. Models with the highest F Scores were selected as part of the presented results.

\begin{table*}[!htbp]
\caption{Comparison of the model performance (for each individual item of MADRS) when output is obtained using hard or soft voting. Each cell contains metrics in the format: Weighted FScore / FScore (Absent) / FScore (Present). Results obtained using the AASS dataset.
Last row corresponds to models trained for Depression Detection (total score $\geq$ 10)} 
\label{table:wfs_aass_hard_soft_sp}
\centering
\begin{tabular}{l|cccc} 
\toprule
& \multicolumn{2}{c}{\textbf{Hard Voting}} & \multicolumn{2}{c}{\textbf{Soft Voting}} \\
\midrule
 \begin{tabular}[c]{@{}c@{}}Item - MADRS\\(AASS dataset) \end{tabular}& \begin{tabular}[c]{@{}c@{}}Spectrogram\\CNN\end{tabular} & \begin{tabular}[c]{@{}c@{}}Spectrogram\\CNN-LSTM\end{tabular} & \begin{tabular}[c]{@{}c@{}}Spectrogram\\CNN\end{tabular} &
 \begin{tabular}[c]{@{}c@{}}Spectrogram\\CNN-LSTM\end{tabular} \\ 
\midrule
(1) Apparent sadness & 0.36/0.32/0.43 & \textbf{0.61/0.71/0.43} & 0.36/0.32/0.43 & \textbf{0.61/0.71/0.43} \\
(2) Reported sadness & 0.62/0.60/0.64 & \textbf{0.71/0.73/0.70} & 0.62/0.60/0.64 & \textbf{0.71/0.73/0.70} \\
(3) Inner tension & \textbf{0.67/0.63}/0.70 &0.65/0.53/\textbf{0.74} & 0.62/0.56/0.67 & 0.65/0.53/\textbf{0.74} \\
(4) Reduced sleep & 0.67/0.67/0.67 & \textbf{0.76/0.76/0.76} & 0.67/0.67/0.67 &\textbf{0.76/0.76/0.76} \\
(5) Reduced appetite & 0.49/0.65/0.00 & \textbf{0.65/0.79/0.22} & 0.49/0.65/0.00 & \textbf{0.65/0.79/0.22} \\
(6) Conc. difficulties &0.41/0.27/\textbf{0.59} &\textbf{0.66/0.72/0.59}& 0.53/0.55/0.50 &\textbf{0.66/0.72/0.59} \\
(7) Lassitude &0.33/\textbf{0.46}/0.13 & \textbf{0.42}/0.40/\textbf{0.45} &0.33/\textbf{0.46}/0.13 & \textbf{0.42}/0.40/\textbf{0.45} \\
(8)  Inability to feel & 0.67/0.72/0.59 &0.52/0.62/0.38 & \textbf{0.79/0.87/0.67} & 0.58/0.73/0.33 \\
(9) Pessimistic thoughts &\textbf{0.76/0.78/0.74} &0.62/0.60/0.64 & 0.67/0.67/0.67 & 0.62/0.60/0.64 \\
(10)  Suicidal thoughts & 0.70/0.74/0.53 &\textbf{0.75/0.86/0.29} & 0.70/0.74/0.53 & \textbf{0.75/0.86/0.29}  \\
\midrule
\begin{tabular}[c]{@{}c@{}}Depression\\Detection\end{tabular} & 0.59/0.61/0.52 & 0.65/0.70/0.54 & 0.61/0.63/0.54 & \textbf{0.66/0.72/0.55} \\
\bottomrule
\end{tabular}
\end{table*}

\section{Experiments}
\label{section:experiments}
The performance of the models is evaluated in terms of F-scores. We provide three F-scores: weighted F-score, F-score of the item being present, and F-score of the item being absent in the given speech sample. We train one model per each item in depression rating scale.
Table \ref{table:wfs_aass_hard_soft_sp} provides the performance of the CNN and CNN-LSTM models when the final decision is obtained using hard or soft voting for each item. The CNN-LSTM models perform better than the CNN models on most of the individual items and also on the task of depression detection. This shows that the local context (temporal information) contained in speech is highly beneficial for prediction of items such as Apparent sadness, Reported sadness, Reduced sleep, Reduced appetite and concentration difficulties. While for items such as Inability to feel and Pessimistic thoughts, CNN-based model performed better than the CNN-LSTM models.

It can also be observed that the use of soft voting compared to hard voting did not significantly effect the predictions on individual items except for the item Inability to feel. But using soft voting slightly improved the performance on the task of depression detection. 
Refer to appendix for partial results on the DAIC-WOZ dataset.

\section{Summary}
In this work, we proposed a two-stage approach for depression detection from speech, motivated by clinician-based depression detection. In other words,  we initially predict the individual items of the depression rating scale before obtaining the final depression prediction. Experimental results show that learning the temporal context of speech is beneficial for predicting most of the items in the depression rating scale. Further, we have shown that the voting scheme used to consolidate the results affects the performance of the models.

\bibliographystyle{unsrtnat}
\bibliography{strings.bib,refs.bib}

\appendix

\section{Appendix}

\subsection{Datasets}
\textbf{Distress Analysis Interview Corpus - Wizard of Oz (DAIC-WOZ)}
The DAIC-WOZ data set \cite{gratch2014distress} consists of interviews held in a Wizard of Oz set up, where the interviewee speaks with a machine that is controlled remotely by a human.
 In order to assess the depression severity of the participants, they are asked to answer the items contained in the self-rated Patient Health Questionnaire-8 (PHQ-8) \cite{phq8}. Where the questionnaire scores represent an answer to the question "In the last 2 weeks how often have you been bothered by any of the following problems?" and the possible answers are: (0) Not at all, (1) Several days, (2) More than half the days, and (3) Nearly every day. In total there are 8 problems in the questionnaire, listed in Table \ref{table:support}, for a maximum score of 24, which indicates severe depression. For this data set we adopt the suggested splits, however, we evaluate on the development split due to the lack of individual PHQ-8 items in the test set, while the training data set is split into train and validation splits. As part of our final setup we are left with 219 recordings, 138 in the train set, 25 in the validation set and 56 in the test set (originally the development set).
 
 \subsection{Autobiographical Adult Speech Samples (AASS)}
This data set contains 131 recordings from 109 speakers. Each of the recordings corresponds to a person speaking about her/his past weeks. During the interview, the interviewees are asked to answer these 3 prompts:
\begin{enumerate}
  \item (Neutral) Tell me how you have been feeling and what you have been up to lately.
  \item (Positive) Think about when you had a positive experience or when something good may have happened to you.
  \item (Negative) Think about when you had a negative experience or when something bad may have happened to you.
\end{enumerate}
Each participant is given a brief introduction for him/her to be ready for the interview, and is also instructed to talk for about 3 minutes in each of the prompts. Some of the participants undertake multiple interviews at different moments to assess their depression levels across time. Afterwards, the recordings are labelled/scored by trained clinical assessors using the Montgomery–Åsberg Depression Rating Scale (MADRS)\cite{montgomery1979madrs}. This scale contains 10 items, listed in Table \ref{table:support}, with scores between 0 to 6, where 6 indicates the highest severity. In extreme cases, the MADRS scale can reach a maximum score of 60. Across our experiments we use a testing split containing 21 recordings, a validation set of 16 recordings and a training set of 94 recordings. Speakers are assigned to one of these sets and do not have any samples in the remaining.

\begin{table}[!htbp]
\caption{Samples on the test split for the AASS and the DAIC-WOZ dataset.}
\label{table:support}
\centering
\begin{tabular}{lcc|lcc} 
\toprule
\multicolumn{3}{c|}{AASS (MADRS)} & \multicolumn{3}{c}{DAIC-WOZ (PHQ-8)} \\
\cmidrule{1-3} \cmidrule{4-6}
Item/Question & \#(Absent) & \#(Present) & Item/Question &\#(Absent) &\#(Present)\\
\midrule
(1) Apparent sadness& 13 & 8 & (1) Little interest& 26 & 30 \\
(2) Reported sadness& 10 & 11 & (2) Feeling down& 26 & 30 \\
(3) Inner tension& 9& 12 & (3) Trouble sleeping& 23 & 33 \\ 
(4) Reduced sleep& 12 & 9 & (4) Feeling tired& 17 & 39 \\
(5) Reduced appetite& 16 & 5 & (5) Poor apetite& 20 & 36 \\
(6) Conc. difficulties& 12&9&(6) Self-disappointment&26&30\\
(7) Lassitude& 13 & 8 & (7) Conc. difficulties& 32 & 24 \\
(8) Inability to feel & 13 & 8 & (8) Restlessness& 40 & 16 \\
(9) Pessimistic thoughts & 10 & 11 & &  \\
(10) Suicidal thoughts & 17 & 4 & & \\ 
\bottomrule
\end{tabular}
\end{table}
\begin{table*}[!tbp]
\caption{Comparison of the model performance (for each individual item of MADRS) when output is obtained using hard or soft voting. Each cell contains metrics in the format: Weighted FScore / FScore (ND) / FScore (D).
} 
\label{table:wfs_aass_hard_soft_egmaps}
\centering
\begin{tabular}{l|cccc} 
\toprule
& \multicolumn{2}{c}{\textbf{Hard Voting}} & \multicolumn{2}{c}{\textbf{Soft Voting}} \\
\midrule
 \begin{tabular}[c]{@{}c@{}}Item - MADRS\\(AASS dataset) \end{tabular}& \begin{tabular}[c]{@{}c@{}}eGeMAPS\\CNN\end{tabular} & \begin{tabular}[c]{@{}c@{}}eGeMAPS\\CNN-LSTM\end{tabular} & \begin{tabular}[c]{@{}c@{}}eGeMAPS\\CNN\end{tabular} &
 \begin{tabular}[c]{@{}c@{}}eGeMAPS\\CNN-LSTM\end{tabular} \\ 
\midrule
(1) Apparent sadness & 0.66/0.74/0.53 & \textbf{0.76/0.81/0.67} & \textbf{0.76/0.81/0.67} &0.62/0.69/0.50 \\
(2) Reported sadness & 0.61/0.67/0.56 &\textbf{0.71}/0.75/\textbf{0.67} & 0.61/0.67/0.56 & 0.69/\textbf{0.77}/0.63 \\
(3) Inner tension & \textbf{0.69/0.57/0.79} &\textbf{0.69/0.57/0.79} &\textbf{0.69/0.57/0.79} & \textbf{0.69/0.57/0.79} \\
(4) Reduced sleep &  0.74/\textbf{0.83}/0.62 & \textbf{0.75}/0.81/\textbf{0.67} &0.74/\textbf{0.83}/0.62 & 0.71/0.77/0.63 \\
(5) Reduced appetite & \textbf{0.62/0.75/0.20} & 0.55/0.73/0.00 & \textbf{0.62/0.75/0.20} &0.55/0.73/0.00 \\
(6) Conc. difficulties &  0.63/0.76/0.46 & \textbf{0.71}/0.75/\textbf{0.67} & 0.69/\textbf{0.79}/0.57 & \textbf{0.71}/0.75/\textbf{0.67} \\
(7) Lassitude &0.57/0.67/0.40 & \textbf{0.67/0.72/0.59} &0.51/0.64/0.29 &0.53/0.55/0.50 \\
(8)  Inability to feel & \textbf{0.80/0.86/0.71} & 0.65/0.76/0.46 & 0.75/0.83/0.62 &0.70/0.79/0.57 \\
(9) Pessimistic thoughts & 0.81/0.82/0.80 & 0.81/0.82/0.80 & \textbf{0.86/0.86/0.86} & 0.81/0.82/0.80 \\
(10)  Suicidal thoughts & \textbf{0.90/0.94/0.75} & 0.62/0.69/0.31 & \textbf{0.90/0.94/0.75} & 0.56/0.58/0.44  \\
\bottomrule
\end{tabular}
\end{table*}
\begin{table*}[!htbp]
\caption{Comparison of the model performance (for each individual item of PHQ-8) when output is obtained using hard or soft voting. Each cell contains metrics in the format: Weighted FScore / FScore (Absent) / FScore (Present). Results obtained using the DAIC-WOZ dataset} 
\label{table:wfs_daic_hard_soft_sp}
\centering
\begin{tabular}{l|cccc} 
\toprule
& \multicolumn{2}{c}{\textbf{Hard Voting}} & \multicolumn{2}{c}{\textbf{Soft Voting}} \\
\midrule
 \begin{tabular}[c]{@{}c@{}}Item - PHQ-8\\(DAIC-WOZ dataset) \end{tabular}& \begin{tabular}[c]{@{}c@{}}Spectrogram\\CNN\end{tabular} & \begin{tabular}[c]{@{}c@{}}Spectrogram\\CNN-LSTM\end{tabular} & \begin{tabular}[c]{@{}c@{}}Spectrogram\\CNN\end{tabular} &
 \begin{tabular}[c]{@{}c@{}}Spectrogram\\CNN-LSTM\end{tabular} \\ 
\midrule
(1) Little interest & 0.51/0.54/0.49 & 0.44/0.47/0.41 & 0.53/\textbf{0.55}/0.52 & \textbf{0.55}/0.38/\textbf{0.69}  \\
(2) Feeling down &0.45/0.35/\textbf{0.55} &0.47/0.38/\textbf{0.55} &\textbf{0.49/0.55}/0.44 &0.47/0.38/\textbf{0.55} \\
(3) Trouble sleeping &0.49/0.27/0.64 &\textbf{0.55/0.47}/0.62 & \textbf{0.55}/0.44/\textbf{0.63} & \textbf{0.55/0.47}/0.62 \\
(4) Feeling tired &\textbf{0.64/0.34/0.77} &0.59/0.22/0.75 &\textbf{0.64/0.34/0.77} &0.59/0.22/0.75 \\
(5) Poor appetite &\textbf{0.55}/0.26/\textbf{0.72} &0.46/\textbf{0.39}/0.49 & 0.51/0.14/0.71 & 0.46/\textbf{0.39}/0.49 \\
(6) Self-disappointment &0.44/0.18/\textbf{0.66} &\textbf{0.54/0.44}/0.63 &0.44/0.18/\textbf{0.66} &0.50/\textbf{0.44}/0.55 \\
(7) Conc. difficulties &0.52/0.57/0.45 &0.52/0.51/0.53 &\textbf{0.59/0.61/0.57} &0.52/0.51/0.53\\
(8) Slow/Agitated &0.61/0.75/0.28 &\textbf{0.68/0.80/0.37} & 0.60/0.77/0.17 & \textbf{0.68/0.80/0.37} \\
\bottomrule
\end{tabular}
\end{table*}

\begin{table*}[!tbp]
\caption{Comparison of the model performance (for each individual item of PHQ-8) when output is obtained using hard or soft voting on DAIC-WOZ dataset. Each cell contains metrics in the format: Weighted FScore / FScore (ND) / FScore (D).
} 
\label{table:wfs_daic_hard_soft_egmaps}
\centering
\begin{tabular}{l|cccc} 
\toprule
& \multicolumn{2}{c}{\textbf{Hard Voting}} & \multicolumn{2}{c}{\textbf{Soft Voting}} \\
\midrule
 \begin{tabular}[c]{@{}c@{}}Item - PHQ-8\\(DAIC-WOZ dataset) \end{tabular}& \begin{tabular}[c]{@{}c@{}}eGeMAPS\\CNN\end{tabular} & \begin{tabular}[c]{@{}c@{}}eGeMAPS\\CNN-LSTM\end{tabular} & \begin{tabular}[c]{@{}c@{}}eGeMAPS\\CNN\end{tabular} &
 \begin{tabular}[c]{@{}c@{}}eGeMAPS\\CNN-LSTM\end{tabular} \\ 
\midrule
(1) Little interest &0.54/0.60/0.49 & \textbf{0.59}/0.60/\textbf{0.58} & 0.53/\textbf{0.63}/0.44 &  \textbf{0.59}/0.61/0.57  \\
(2) Feeling down & \textbf{0.61/0.51/0.70} &0.43/0.38/0.47 & 0.49/0.34/0.62 & 0.48/0.47/0.49  \\
(3)  Trouble sleeping &0.45/0.22/0.62 &\textbf{0.51}/0.32/\textbf{0.65} & \textbf{0.51/0.37}/0.61 & 0.50/0.31/0.63 \\
(4) Feeling tired & 0.56/0.09/0.76 &0.53/\textbf{0.14}/0.70 &0.56/0.09/0.76 &\textbf{0.58}/0.10/\textbf{0.79} \\
(5) Poor apetite &0.48/0.29/0.59 &\textbf{0.54/0.43/0.61} & 0.46/0.25/0.58 & 0.47/0.42/0.50 \\
(6) Self-disappointment &\textbf{0.46/0.40}/0.52 & 0.42/0.13/\textbf{0.67} &\textbf{0.46/0.40}/0.52 & 0.42/0.13/\textbf{0.67} \\
(7) Conc. difficulties &\textbf{0.60/0.69}/0.48 & 0.57/0.68/0.44 &0.55/0.60/\textbf{0.49} &0.55/0.67/0.40 \\
(8) Slow/Agitated &0.61/0.75/0.28 &\textbf{0.68/0.80/0.37} &0.64/0.78/0.30 &0.60/0.75/0.22 \\
\bottomrule
\end{tabular}
\end{table*}

\begin{table*}[!htbp]
\caption{Comparison of the model performance (for each individual item of MADRS) obtained using multi-task and single-task models. Results are presented in terms of Weighted FScore. Results obtained using the AASS dataset
} 
\label{table:wfs_aass_multitask_sp}
\centering
\begin{tabular}{l|cccccccc} 
\toprule
& \multicolumn{4}{c}{\textbf{Hard Voting}} & \multicolumn{4}{c}{\textbf{Soft Voting}} \\
& \multicolumn{2}{c}{\textbf{Single-task}} & \multicolumn{2}{c}{\textbf{Multi-task}} & \multicolumn{2}{c}{\textbf{Single-task}} & \multicolumn{2}{c}{\textbf{Multi-task}} \\
\midrule
 \begin{tabular}[c]{@{}c@{}}Item - MADRS\\(AASS dataset) \end{tabular}& \begin{tabular}[c]{@{}c@{}}Spec\\CNN\end{tabular} & \begin{tabular}[c]{@{}c@{}}Spec\\C-L\end{tabular} & \begin{tabular}[c]{@{}c@{}}Spec\\CNN\end{tabular} &
 \begin{tabular}[c]{@{}c@{}}Spec\\C-L\end{tabular} &
\begin{tabular}[c]{@{}c@{}}Spec\\CNN\end{tabular} & \begin{tabular}[c]{@{}c@{}}Spec\\C-L\end{tabular} & \begin{tabular}[c]{@{}c@{}}Spec\\CNN\end{tabular} &
 \begin{tabular}[c]{@{}c@{}}Spec\\C-L\end{tabular} \\ 
\midrule
(1) Apparent sadness & 0.36 & 0.61 & 0.52 & \textbf{0.69} & 0.36 & 0.61 & 0.52 & \textbf{0.69} \\
(2) Reported sadness & 0.62 & \textbf{0.71} & 0.59 & 0.62 & 0.62 & \textbf{0.71} & 0.59 & 0.62 \\
(3) Inner tension & \textbf{0.67} & 0.65 & 0.62 & 0.49 & 0.62 & 0.65 & 0.62 & 0.45 \\
(4) Reduced sleep & 0.67 & \textbf{0.76} & 0.67 & 0.62 & 0.67 & \textbf{0.76} & 0.67 & 0.62 \\
(5) Reduced appetite & 0.49 & \textbf{0.65} & 0.58 & 0.58 & 0.49 & \textbf{0.65} & 0.58 & 0.58 \\
(6) Conc. difficulties & 0.41 & \textbf{0.66} & 0.57 & 0.53 & 0.53 & \textbf{0.66} & 0.57 & 0.53 \\
(7) Lassitude & 0.33 & 0.42 & \textbf{0.52} & 0.44 & 0.33 & 0.42 & \textbf{0.52} & 0.44 \\
(8) Inability to feel & 0.67 & 0.52 & 0.48 & 0.57 & \textbf{0.79} & 0.58 & 0.48 & 0.51 \\
(9) Pessimistic thoughts & \textbf{0.76} & 0.62 & \textbf{0.76} & 0.71 & 0.67 & 0.62 & \textbf{0.76} & 0.71 \\
(10) Suicidal thoughts & 0.70 & \textbf{0.75} & 0.68 & 0.70 & 0.70 & \textbf{0.75} & 0.68 & 0.70 
\end{tabular}
\end{table*}
\begin{table*}[!htbp]
\caption{Comparison of the model performance (for each individual item of MADRS) obtained using multi-task and single-task models. Results are presented in terms of Weighted FScore. Results obtained using the AASS dataset
} 
\label{table:wfs_aass_multitask_egemaps}
\centering
\begin{tabular}{l|cccccccc} 
\toprule
& \multicolumn{4}{c}{\textbf{Hard Voting}} & \multicolumn{4}{c}{\textbf{Soft Voting}} \\
& \multicolumn{2}{c}{\textbf{Single-task}} & \multicolumn{2}{c}{\textbf{Multi-task}} & \multicolumn{2}{c}{\textbf{Single-task}} & \multicolumn{2}{c}{\textbf{Multi-task}} \\
\midrule
 \begin{tabular}[c]{@{}c@{}}Item - MADRS\\(AASS dataset) \end{tabular}& \begin{tabular}[c]{@{}c@{}}eGeM\\CNN\end{tabular} & \begin{tabular}[c]{@{}c@{}}Spec\\C-L\end{tabular} & \begin{tabular}[c]{@{}c@{}}eGeM\\CNN\end{tabular} &
 \begin{tabular}[c]{@{}c@{}}eGeM\\C-L\end{tabular} &
\begin{tabular}[c]{@{}c@{}}eGeM\\CNN\end{tabular} & \begin{tabular}[c]{@{}c@{}}Spec\\C-L\end{tabular} & \begin{tabular}[c]{@{}c@{}}eGeM\\CNN\end{tabular} &
 \begin{tabular}[c]{@{}c@{}}eGeM\\C-L\end{tabular} \\ 
\midrule
(1) Apparent sadness & 0.66 & 0.76 & 0.70 & 0.75 & 0.76 & 0.62 & \textbf{0.77} & 0.66 \\
(2) Reported sadness & 0.61 & 0.71 & 0.71 & 0.76 & 0.61 & 0.69 & 0.71 & \textbf{0.81} \\
(3) Inner tension & 0.69 & 0.69 & 0.65 & \textbf{0.75} & 0.69 & 0.69 & 0.65 & 0.69 \\
(4) Reduced sleep & 0.74 & 0.75 & \textbf{0.86} & 0.82 & 0.74 & 0.71 & \textbf{0.86} & 0.76 \\
(5) Reduced appetite & 0.62 & 0.55 & \textbf{0.72} & 0.69 & 0.62 & 0.55 & \textbf{0.72} & 0.69 \\
(6) Conc. difficulties & 0.63 & 0.71 & 0.66 & 0.71 & 0.69 & 0.71 & 0.65 & \textbf{0.81} \\
(7) Lassitude & 0.57 & 0.67 & \textbf{0.71} & 0.57 & 0.51 & 0.53 & \textbf{0.71} & 0.58 \\
(8) Inability to feel & 0.80 & 0.65 & 0.80 & 0.80 & 0.75 & 0.70 & 0.80 & \textbf{0.86} \\
(9) Pessimistic thoughts & 0.81 & 0.81 & \textbf{0.86} & \textbf{0.86} & \textbf{0.86} & 0.81 & \textbf{0.86} & \textbf{0.86} \\
(10) Suicidal thoughts & 0.90 & 0.62 & \textbf{0.71} & 0.75 & \textbf{0.91} & 0.56 & \textbf{0.71} & 0.77 \\
\bottomrule
\end{tabular}
\end{table*}
\begin{table*}[!htbp]
\caption{Comparison of the model performance (for each individual item of MADRS) obtained using multi-task and single-task models. Results are presented in terms of Weighted FScore. Results obtained using the DAIC-WOZ dataset
} 
\label{table:wfs_daic_multitask_sp}
\centering
\begin{tabular}{l|cccccccc} 
\toprule
& \multicolumn{4}{c}{\textbf{Hard Voting}} & \multicolumn{4}{c}{\textbf{Soft Voting}} \\
& \multicolumn{2}{c}{\textbf{Single-task}} & \multicolumn{2}{c}{\textbf{Multi-task}} & \multicolumn{2}{c}{\textbf{Single-task}} & \multicolumn{2}{c}{\textbf{Multi-task}} \\
\midrule
 \begin{tabular}[c]{@{}c@{}}Item - PHQ-8\\(DAIC-WOZ dataset) \end{tabular}& \begin{tabular}[c]{@{}c@{}}Spec\\CNN\end{tabular} & \begin{tabular}[c]{@{}c@{}}Spec\\C-L\end{tabular} & \begin{tabular}[c]{@{}c@{}}Spec\\CNN\end{tabular} &
 \begin{tabular}[c]{@{}c@{}}Spec\\C-L\end{tabular} &
\begin{tabular}[c]{@{}c@{}}Spec\\CNN\end{tabular} & \begin{tabular}[c]{@{}c@{}}Spec\\C-L\end{tabular} & \begin{tabular}[c]{@{}c@{}}Spec\\CNN\end{tabular} &
 \begin{tabular}[c]{@{}c@{}}Spec\\C-L\end{tabular} \\ 
\midrule
(1) Little Interest & 0.51 & 0.44 & \textbf{0.55} & 0.48 & 0.53 & \textbf{0.55} & \textbf{0.55} & 0.48 \\
(2) Feeling down & 0.45 & 0.47 & 0.48 & 0.39 & \textbf{0.49} & 0.47 & 0.48 & 0.41 \\
(3) Trouble sleeping & 0.49 & \textbf{0.55} & \textbf{0.55} & \textbf{0.55} & \textbf{0.55} & \textbf{0.55} & \textbf{0.55} & \textbf{0.55} \\
(4) Felling tired & 0.64 & 0.59 & \textbf{0.67} & 0.54 & 0.64 & 0.59 & \textbf{0.67} & \textbf{0.54} \\
(5) Poor apetite & 0.55 & 0.46 & \textbf{0.58} & \textbf{0.49} & 0.51 & 0.46 & \textbf{0.58} & 0.51 \\
(6) Self-disappointment & 0.44 & \textbf{0.54} & 0.53 & \textbf{0.33} & 0.44 & 0.5 & \textbf{0.54} & \textbf{0.33} \\
(7) Conc. difficulties & 0.52 & 0.52 & 0.49 & 0.52 & \textbf{0.59} & 0.52 & 0.49 & 0.52 \\
(8) Slow/Agitated & 0.61 & \textbf{0.68} & 0.53 & 0.61 & 0.6 & \textbf{0.68} & \textbf{0.53} & 0.63 
\end{tabular}
\end{table*}

\end{document}